# RADIOSEISMOLOGY AS A NEW METHOD OF INVESTIGATIONS OF METEOR STREAMS ON THE MOON AND PLANETS

**Berezhnoi A.A. [(1)], Bervalds E. [(2)], Khavroshkin O.B. [(3)], Ozolins G. [(2)]**

*(1) Sternberg Astronomical Institute, Universitetskij pr. 13, 119899 Moscow, Russia, Email: ber@sai.msu.ru*
*(2) Ventspils International Radio Astronomy Center, Akademijas laukums 1, LV-1050 Riga, Latvia, Email: berv@latnet.lv*

*(2) Ventspils International Radio Astronomy Center, Akademijas laukums 1, LV-1050 Riga, Latvia, Email: ozo@acad.latnet.lv*
*(3) Institute of Earth Physics, B. Gruzinskaya 10, 123810 Moscow, Russia, Email: khavole@uipe-ras.scgis.ru*

## 1. ABSTRACT

Radioseismology is based on registration and interpretation of radio emission of seismic origin. Such radioseismic processes occur on the Moon, planets, and asteroids. Non-thermal radio emission of the Moon caused by rock fracturing, seismic activity, and thermal cracking of the regolith was detected during observations of the Moon at the 64 m Kaliazin radio telescope at 13 and 21 cm on July 30 – August 2, 1999. We observed the Moon with the 32 m Ventspils antenna at 25 mm on November 16 - 18, 2000. During the morning of November 17 we registered significant quasiperiodic oscillations of the lunar radio emission starting near 1:44 UT and continuing until the end of observations at 7:17 UT. Oscillations were also registered on November 18 starting near 2:28 UT. Intensive oscillations were registered until about 7:00 UT with bottom to peak heights of 5-10 K. The time of maxima of oscillations does not contradict theoretical predictions about the existence of three maxima of the Leonid meteor shower on the Moon. Amplitudes of oscillations were equal to 1-2 K before and after the time of Leonid's maxima. These results can be explained as the detection of lunar radio emission of seismic origin caused by meteoroid impacts. The implications of the radioseismic method for determination of the intensity of meteor showers on the Moon and planets and the internal structure of the Moon are described.

## 2. INTRODUCTION

There are some sources of radio emission on the Moon: the thermal radiation, reflected from the Moon solar radiation and radiation of our Galaxy, and the radio emission of seismic origin. The main component of the lunar radio emission is the thermal radiation. Investigations of the lunar thermal radio emission can estimate density and electric properties of the lunar regolith, temperature regime and thermal gradient of surface layers (see for example [1]). Reflected solar and galactic radiation are essential only for observations of the Moon at $\lambda > 50$ cm. However during intensive solar flares it is possible to detect additional lunar radio emission caused by such flares at $\lambda > 5$ cm.

Radioseismology is based on registration and interpretation of electromagnetic radiation of seismic origin. The frequency of such radiation before earthquakes [2] and during the rock fracture experiments [3] lies in the range from the frequency of kHz radio to the frequency of optical and soft X-ray radiation. There are not articles about the detection of radio emission during rock fracture at 100 MHz - $10^{14}$ Hz. The probable three models of transformation of mechanical stress into electromagnetic radiation are: 1) the formation of new microcracks; 2) charges arising at the peaks of existing cracks forming under the action of increasing load; 3) the piezoeffect. Radio emission associated with quartz-bearing rock fracture was detected at 1-2 MHz [4]. The nature of such electromagnetic emission has not been investigated very well. Such radioseismic processes occur on the Moon and airless planets and asteroids. Non-thermal radio emission of the Moon ($\nu = 1 - 5$ kHz) caused by rock fracturing, seismic activity, and thermal cracking during lunar transient phenomena is discussed in [5]. The possibility of the detection of collisions of cosmic rays with energies from $10^{20}$ to $10^{23}$ eV with the Moon during lunar radio observations at $\lambda = 1 - 100$ m was discussed in [6].

## 3. LUNAR OBSERVATIONS AFTER THE LUNAR PROSPECTOR IMPACT AT 13 AND 21 CM

The collision between the American spacecraft Lunar Prospector and the Moon was at 9:52 UT on July 31, 1999. Due to singularities of the lunar lithosphere the duration of seismic effect accompanied this collision may be equal to some hours. Lunar non-thermal radio emission was detected during observations of the Moon at the 64-m Kaliazin radio telescope at 13 and 21 cm on July30 - August 2, 1999 [7].

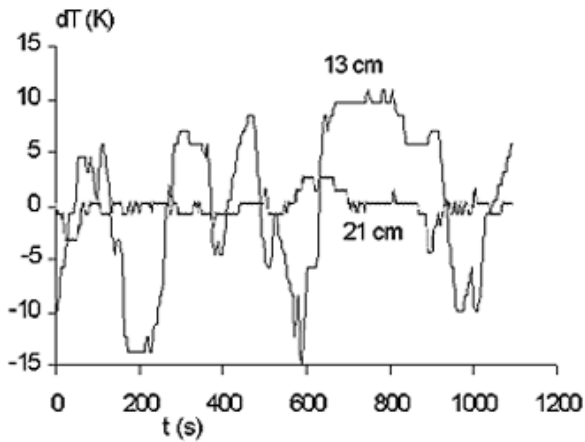

Fig. 1. Variations of the radio flux (in K) of a seismically active region of the Moon (30W, 5S) at 13 and 21 cm versus time. Starting point is July 31, 1999; 22:38 UT.

The mean amplitude of variations of the lunar radio flux at 13 cm was 4, 8, 10, 3 K (and at 13 cm - 2, 2, 4, 3 K) on July 29/30, July 30/31, July 31/August 1, August 1/2 respectively. The correlation between variations of the total lunar radio flux at 13 and 21 cm was detected also [8]. The correlation coefficient was equal to 0.5. This fact can be explained as a similar nature of the non-thermal radio emission at both frequencies. Oscillations with periods of about 20 s and 1 - 11 minutes were registered during observations of the Moon at 13 and 21 cm (see fig. 1).
Short periods (60 s or less) can be caused by the atmospheric turbulence. Namely periods at 13 cm were equal to 1.5, 1.5, 1.7, 2.8, 3.1, 3.5, 4.2, 6.7, 8.5, and 11.7 minutes. Periods at 21 cm were equal to 1.5, 2.8, and 6.8 minutes. Such periods are close to theoretical predictions of periods of free oscillations of the Moon [9]. Meteorite bombardment and moonquakes excite oscillations of the Moon, which lead to generation of radioseismic emission. Accurate determination of periods of free oscillations of the Moon can increase our knowledge about the internal structure of the Moon.
Let us compare the kinetic energy of the Lunar Prospector impact (M ~ 200 kg, V ~ 1.7 km/s, E ~ $3*10^8$ J) and meteorite bombardment. The current estimation of the micrometeorite flux into Earth and the Moon is $10^{-7}$ kg/year*m$^2$ [10]. So the micrometeorite flux into the Moon is about $5*10^3$ kg per day. At the mean speed of micrometeorites of about 20 km/s the power of micrometeorite bombardment of the Moon is about $10^7$ Wt. Comparing the kinetic energy of the Lunar Prospector impact with the intensity of micrometeorite bombardment of the Moon we can conclude that the Lunar Prospector impact can not explain increasing of the non-thermal lunar radio emission at 13 cm on July 31, 1999.

## 4. THE MOON AT 25 MM DURING THE LEONID 2000 METEOR SHOWER

For determination of dependence of the intensity of the non-thermal lunar radio emission from micrometeorite bombardment we conducted observations of the Moon during the Leonid 2000 meteor shower. The Leonid meteor stream was the strongest meteor stream crossing the Earth orbit in 1999 and 2000. Optical flashes on the dark side of the Moon caused by Leonid meteoroid impacts were detected on November 1999 [11]. We observed the Moon on November 16 - 18 with the 32 m antenna of the Ventspils International Radio Astronomy Center at λ = 25 mm [12]. The half-power beam width was 3.5 arcminutes. The bandwidth was 44 MHz, and the output time constant was 1 s. We could not exactly track the antenna with the velocity of the Moon, and the observable region lagged behind and during 30 minutes of observation cycle the beam draw a 15 arcminutes long trip on the lunar surface in direction to Mare Serenitatis. Then the next half - hour scanning repeated starting from a seismic active lunar region (30W, 5S).

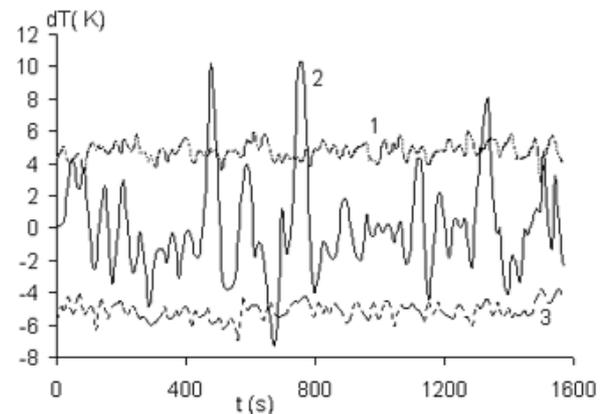

Fig. 2. Variations of the lunar radio flux (in K) at 25 mm during the Leonid 2000 meteor shower versus time. Starting points are Nov. 17, 22:17 UT (curve1), Nov. 18, 4:31 UT (curve 2), Nov. 18, 7:32 UT (curve 3).

During the morning of November 17 we registered significant quasiperiodic oscillations of the lunar radio emission starting near 1:44 UT and continuing until the end of observations at 7:17 UT. There were at least two isolated events with amplitudes exceeding 10 K at 3:12 UT and 4:33 UT. The maximal amplitudes of observed variations were registered on November 18, 4-5 UT also (see fig. 2). Intensive oscillations were registered until about 7 UT with bottom to peak heights of some K, sometimes up to 10 K. The periods were equal to 1.8, 2.2, 3.0, 3.6, 4.4 minutes. Disagreement between values of periods at 25 mm and 13, 21 cm can be explained by different observational method of lunar radio observations (Kaliazin radio telescope can follow to the Moon, Ventspils radio telescope can not follow to the Moon). The time of maxima of oscillations does not contradict to theoretical results about the existence at least first and third maxima from three predicted maxima of the Leonid shower on the Moon [13] (November 17, 4:58 UT (the predicted zenith rate ZHR is about 1000 per hour); November 18, 0:52 (ZHR ~ 10) and 5:01 UT (ZHR ~ 50)). Amplitudes of oscillations were equal to 1-2 K before and after the time of maxima of the Leonid meteor shower. The correlation between variations of the lunar radio flux from same lunar regions at different moments of observations is absent. It is an evidence of weak dependence of the intensity of non-thermal radio emission from the mineral composition of the lunar regolith.

We hope that our future observations of the Moon can solve the following unresolved problems: determination of the spectral index and the total power of the non-thermal radio emission, estimation of intensity of such emission as a function of mass and speed of impactors. What source (micrometeorite bombardment, moonquakes or thermal cracking) is the main source of the non-thermal emission at different wavelengths? For solving of these problems the organisation of continuum radio observations of the Moon at $\lambda = 1$ mm – 1 m is desirable.

## 5. CONCLUSIONS

Our results of observations of the Moon at 25 mm during the Leonid meteor shower can be explained as the detection of lunar radio emission of seismic origin caused by meteoroid impacts.

Oscillations of the lunar radio flux can be explained as a result of excitation of free oscillations of the Moon and the transformation of seismic energy to electromagnetic radiation.

Radio observations of the Moon can help us to determine the intensity of meteor showers on the Moon and the structure of meteoroid streams.

In the future the detection of radio emission of seismic origin from Earth-type planets is possible.